\documentstyle[epsf]{lamuphys}

\newcommand{\beq}{\begin{equation}}
\newcommand{\eeq}{\end{equation}}
\newcommand{\bqa}{\begin{eqnarray}}
\newcommand{\eqa}{\end{eqnarray}}

\newcommand{\eq}[1]{Eq.~(\ref{#1})}

\newcommand{\fig}[1]{Fig.~\ref{#1}}

\newcommand{\ie}{{\frenchspacing{\em i.\hspace{0.04cm}e.}}}

\newcommand{\dof}{{\frenchspacing{\em d.o.f.}}}

\newcommand{\emu}{{\vec e}_{\mu}}
\newcommand{\enu}{{\vec e}_{\nu}}
\newcommand{\tr}{\mbox{Tr}}
\newcommand{\tauint}{\tau_{\mbox{\footnotesize int}}}
\makeatletter
\let\chapter\hid@chapter
\makeatother
\begin{document}
\pagenumbering{arabic}
\title{The Hybrid Monte Carlo Algorithm for Quantum Chromodynamics}

\author{Thomas\,Lippert}

\institute{Department of Physics, University of Wuppertal, 
D-42097 Wuppertal, Germany}

\maketitle

\begin{abstract}
  The Hybrid Monte Carlo (HMC) algorithm currently is the favorite
  scheme to simulate quantum chromodynamics including dynamical
  fermions. In this talk---which is intended for a non-expert
  audience---I want to bring together methodical and practical aspects
  of the HMC for full QCD simulations.  I will comment on its merits
  and shortcomings, touch recent improvements and try to forecast its
  efficiency and r\^ole in future full QCD simulations.
\end{abstract}

\section{Introduction}
The Hybrid Monte Carlo algorithm \cite{DUANE87} is---for the
present---a culmination in the development of practical simulation
algorithms for full quantum chromodynamics (QCD) on the lattice.  QCD
is the theory of the strong interaction. In principle, QCD can
describe the binding of quarks by gluons, forming the hadrons with
their masses, as well as other hadronic properties.  As QCD cannot be
evaluated satisfactorily by perturbative methods, one has to recourse
to {\em non-perturbative} stochastic simulations of the quark and
gluon fields on a discrete 4-dimensional space-time lattice
\cite{CREUTZ}.  In analogy to simulations in statistical mechanics, in
a Markov chain, a canonical ensemble of field configurations is
generated by suitable Monte Carlo algorithms.  As far as full QCD
lattice simulations are concerned, the HMC algorithm is the method of
choice as it comprises several important advantages:
\begin{itemize}
\item The evolution of the gluon fields through phase space is carried
  out simultaneously for all \dof, as in a {\em molecular dynamics}
  scheme, using the leap-frog algorithm or higher order symplectic
  integrators.
\item {\em Dynamical fermion loops}, represented in the path-integral
  in form of a determinant of a huge matrix of dimension $O(10^7)$
  elements, \ie\ a highly non-local object that is not directly
  computable, can be included by means of a stochastic representation
  of the fermionic determinant.  This approach amounts to the solution
  of a huge system of linear equations of rank $O(10^7)$ that can be
  solved efficiently with modern iteration algorithms, so-called
  Krylov-subspace methods \cite{FROMMER94,FISCHER96}.
\item As a consequence, the computational complexity of HMC is a
  number $O(V)$, \ie, one complete sweep (update of all V \dof)
  requires $O(V)$ operations, as it is the case for Monte Carlo
  simulation algorithms of local problems.
\item HMC is {\em exact}, \ie\ systematic errors arising from finite
  time steps in the molecular dynamics are eliminated by a {\em
    global} Monte Carlo decision.
\item HMC is {\em ergodic} due to {\em Langevin}-like stochastic
  elements in the field update.
\item HMC shows surprisingly short {\em autocorrelation times}, as
  recently demonstrated \cite{SESAM97AUTO}. The autocorrelation
  determines the statistical significance of physical results computed
  from the generated ensemble of configurations.
\item HMC can be fully {\em parallelized}, a property that is essential for
  efficient simulations on high speed parallel systems.
\item HMC is computation dominated, in contrast to memory intensive
  alternative methods \cite{LUESCHER94,SLAVNOV95}. Future high
  performance SIMD (single addressing multiple data) systems
  presumably are memory bounded.
\end{itemize}
In view of these properties, it is no surprise that all large scale
lattice QCD simulations including dynamical Wilson fermions as of
today are based on the HMC algorithm.  Nevertheless, dynamical fermion
simulations are still in their infancy.  The computational demands of
full QCD are huge and increase extremely if one approaches the {\em
  chiral limit} of small quark mass, \ie\ the physically relevant mass
regime of the light {\sf u} and {\sf d} quarks.  The central point is
the solution of the linear system of equations by iterative methods.
The iterative solver, however, becomes increasingly inefficient for
small quark mass.  We hope that these demands can be satisfied by
parallel systems of the upcoming tera-computer class
\cite{SCHILLING97}.

The HMC algorithm is a general {\em global} Monte Carlo procedure that
can evolve all \dof\ of the system at the same instance in time.
Therefore it is so useful for QCD where due to the inverse of the
local fermion matrix in the stochastic representation of the fermionic
determinant the gauge fields must be updated all at once to achieve
$O(V)$ complexity.  The trick is to stay close to the surface of
constant Hamiltonian in phase space, in order to achieve a large
acceptance rate in the global Monte Carlo step.

HMC can be applied in a variety of other fields.  A promising novel
idea is the merging of HMC with the multi-canonical algorithm
\cite{BERG} which is only parallelizable within global update schemes.
The parallel multi-canonical procedure, can be applied at the
(first-order) phase transitions of compact QED and Higgs-Yukawa model.
Another example is the Fourier accelerated simulation of polymer
chains as discussed in Anders Irb\"ack's contribution to these
proceedings, where HMC well meets the non-local features of Fourier
acceleration leading to a {\em multi-scale} update process.

The outline of this talk is as follows: In section \ref{SEC:ELEMENTS},
a minimal set of elements and notions from QCD, necessary for the
following, is introduced.  In section \ref{SEC:HMC}, the algorithmic
ingredients and computational steps of HMC are described.  In section
\ref{SEC:COMPLEX}, I try to evaluate the computational complexity of
HMC and suggest a scaling rule of the required CPU-time for vanishing
Wilson quark mass.  Using this rule, I try to give a prognosis as to
the r\^ole of HMC in future full QCD simulations in relation to
alternative update schemes.

\section{Elements of Lattice QCD\label{SEC:ELEMENTS}}
I intend to give a pedagogical introduction into the HMC evaluation of
QCD in analogy to Monte Carlo simulations of statistical systems.
Therefore, I avoid to focus on details.  I directly introduce the
physical elements on the discrete lattice that are of importance for
the HMC simulation. For the following, we do not need to discuss their
parentage and relation to continuum physics in detail.

QCD is a constituent element of the standard model of elementary
particle physics. Six quarks, the flavors up, down, strange, charm,
bottom, and top interact via gluons.  In 4-dimensional space-time, the
fields associated with the quarks, $\psi^{\alpha}_{a}(\vec x)$ have
four Dirac components, $\alpha=1,\dots,4$, and three color components,
$a=1,\dots,3$. The `color' degree of freedom is the characteristic
property reflecting the non-abelian structure of QCD as a gauge
theory. This structure is based on local SU(3) gauge group
transformations acting on the color index.

The gluon field $A_{\mu}^{a}(\vec x)$ consists of four Lorentz-vector
components, $\mu=1,\dots,4$.  Each component carries an index $a$
running from $1$ to $8$. It refers to the components of the eight
gluon field in the basis of the eight generators $\lambda_a$ of the
group SU(3). The eight $3\times 3$ matrices $\lambda_a/2$ are
traceless and hermitean defining the algebra of SU(3) by $[
\frac{\lambda_i}{2},\frac{\lambda_j}{2} ]
=i\,f_{ijk}\,\frac{\lambda_k}{2}.$\footnote{For the explicit structure
  constants $f_{ijk}$ and the generators $\lambda_i/2$ see
  \cite{CHENGLI}.}

\begin{figure}
\epsfxsize=.5\textwidth\epsfbox{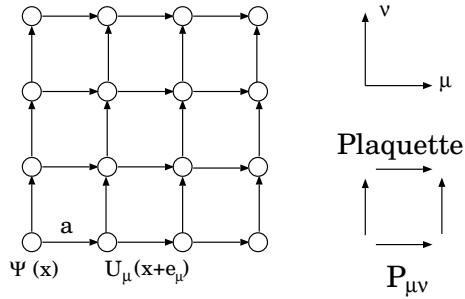}
\caption{2-dimensional projection of the 4-dimensional euclidean
  space-time lattice.\label{FIG:LATTICE}}
\end{figure}

On the lattice, the quark fields $\psi_{\vec n}$ are considered as
approximations to the continuum fields $\psi(\vec x)$, with $\vec x =
a\vec n$, $\vec n\in {\bf N^4}$ (All lattice quantities are taken
dimensionless in the following.).  As shown in \fig{FIG:LATTICE}, they
`live' on the sites.  Their fermionic nature is expressed by
anti-commutators,
\begin{equation}
[\psi^{\alpha}_{\vec n}, 
\psi^{\beta}_{\vec m}]_{+}=
[\psi^{\dagger\alpha}_{\vec n}, 
\psi^{\beta}_{\vec m}]_{+}=
[\psi^{\dagger\alpha}_{\vec n}, 
\psi^{\dagger\beta}_{\vec m}]_{+}=0,
\end{equation}
characterizing the quark fields as Grassmann variables. 
The gluon fields in the 4-dimensional discretized world are represented as
bi-local objects, the so-called links $U_{\mu}(\vec n)$.  They are the
bonds between site $\vec n$ and site $\vec n+\emu$, with $\emu$ being the
unit vector in direction $\mu$. Unlike the continuum gluon field, the gluon
in discrete space is $\in$ SU(3).  $U_{\mu}(\vec n)$ is a discrete
approximation to the {parallel transporter} known from continuum QCD,
$U(x,y)=\exp{(ig_s\int_{\vec{x}}^{\vec y}d{x'}^{\mu}\,A_{\mu}^{a}(\vec
  x')\lambda_a/2)}$, with $g_{s}$ being the strong coupling constant.

QCD is defined via the action $S=S_g+S_f$ that consists of the pure
gluonic part and the fermionic action.  The latter accounts for the
quark gluon interaction and the fermion mass term.  Taking the link
elements from above one can construct a simple quantity, the plaquette
$P_{\mu\nu}$, see \fig{FIG:LATTICE}:
\begin{equation}
P_{\mu\nu}(\vec n)=
U_{\mu}(\vec n)
U_{\nu}(\vec n+\emu)
U^{\dagger}_{\mu}(\vec n+\enu)
U^{\dagger}_{\nu}(\vec n). 
\end{equation}
The Wilson gauge action is defined by means of the plaquette:
\begin{equation}
\beta S_g=\frac{6}{{g_s}^2}\sum_{\vec n,\mu,\nu}\left[
1-\frac{1}{2}\tr(
P_{\mu\nu}(\vec n)+P^{\dagger}_{\mu\nu}(\vec n))
\right].
\end{equation}
In the limit of vanishing lattice spacing, one can recover the
continuum version of the gauge action, $-\int d^4\vec x\frac{1}{4}
F_{\mu\nu}(\vec x)F^{\mu\nu}(\vec x)$.  The deviation from the
continuum action due to the finite lattice spacing $a$ is of $O(a^2)$.

The discrete version of the fermionic action cannot be constructed by
a simple differencing scheme, as it would correspond to 16 fermions
instead of 1 fermion in the continuum limit.  One method to get rid of
the doublers is the addition of a second order
derivative term, 
$(\psi_{\vec n+\emu}- 2\psi_{\vec n}-\psi_{\vec
  x-\emu})/2$, to the standard first order derivative
$\gamma^{\mu}\partial_{\mu}\psi(\vec x)\rightarrow
\gamma^{\mu}(\psi_{\vec n+\emu} - \psi_{\vec n-\emu})/2$.  This scheme
is called Wilson fermion discretization. The fermionic action can be
written as a bilinear form, $S_f=\bar\psi_{\vec n}M_{\vec n\vec
  y}\psi_{\vec m}$, with the Wilson matrix $M$,
\begin{equation}
  {M}_{\vec n\vec m} = \delta_{\vec n\vec m}-{\kappa}\sum_{\mu=1}^4
    (1-{
    \gamma}_{\mu}){U}_{\mu}(\vec n)\, \delta_{\vec n,\vec m-\emu}
    (1+{
    \gamma}_{\mu}){U}^{\dagger}_{\mu}(\vec n-\emu) \delta_{\vec n,\vec
    m+{\emu}}.
\end{equation}

The stochastic simulation of QCD starts from the analogy of the
path\-integ\-ral---the quantization prescription---to a partition sum
as known from statistical mechanics.  As it is oscillating, it would
be useless for stochastic evaluation.  The appropriate framework for
stochastic simulation of QCD is that of Euclidean field theory.
Therefore, one performs a rotation of the time direction $t\rightarrow
i\tau$. The ensuing effect is a transformation of the Minkowski
metrics into a Euclidean metrics, while a positive definite Boltzmann
weight $\exp(-\beta S_g)$ is achieved.  This form of the
path-integral, \ie\ the partition function, is well known from
statistical mechanics:
\begin{equation}
Z= \int\left(\prod_{\vec n,{\mu}}[d{U_{\mu}}(
\vec    x)][d{\bar\psi_{\vec n}}][d\psi_{\vec n}]\right) e^{-\beta
      S_g-{S}_{f}}.
\end{equation}
It is important for the following that one can integrate out the
bilinear $S_f$ over the Grassmann fermion fields.  As a result, we
acquire the determinant of the fermionic matrix:
\begin{equation}
  Z=\int\prod_{\vec n,{\mu}}[d{U_{\mu}}(
\vec n)] \det(M[U])
e^{-{\beta}{S}_{g}}.
\label{PATHINT}
\end{equation}

\section{Hybrid Monte Carlo\label{SEC:HMC}}
The Euclidean path-integral, \eq{PATHINT}, can in principle be
evaluated by Monte Carlo techniques. We see that the fermionic fields
do not appear in $Z$ after the integ\-ration\footnote{Similarly, one
  can perform the computation of any correlation function of
  $\bar\psi$ and $\psi$, leading to products of the quark propagator,
  \ie\ the inverse of $M^{-1}$.}. Hence, it suffices to generate a
representative ensemble of fields $\{ U_i \}$, $i=1,\dots,N$, and
subsequently, to compute any observable along with the statistical
error according to
\begin{equation}
  \langle O\rangle=\frac{1}{N} \sum_{i=1}^{N}
  O_{i}[{U}_{i}]\quad\mbox{and}\quad
\sigma^{2}_{O}
=\frac{2\tauint}{N}\left(
\frac{1}{N}
\sum_{i=1}^{N}
|O_{i}[{ U}_{i}]|^{2}-\langle O\rangle^{2}\right).
\end{equation}
The integrated autocorrelation time $\tauint$ reflects the fact that
the members of the ensemble are generated by importance sampling in a
Markov chain. Therefore, a given configuration is correlated with its
predecessors, and the actual statistical error of a result is
increased compared to the naive standard deviation.  The length of the
autocorrelation time is a crucial quantity for the efficiency of a
simulation algorithm.

\subsection{$O(V)$ Algorithms for full QCD}

If we want to generate a series of field configurations $U_1$, $ U_2
$, $ U_3 $\dots in a Markov process, besides the requirement for
ergodicity, it is sufficient to fulfill the condition of detailed
balance to yield configurations according to a canonical probability
distribution:
\begin{equation}
e^{-S}P(U\rightarrow U')=e^{-S'}P(U'\rightarrow U).
\label{DETAILED}
\end{equation}
$P(U\rightarrow U')$ is the probability to arrive at configuration
$U'$ starting out from $U$.  Let us for the moment forget about
$\det(M[U])$, \ie, we set $\det(M)$ equal to 1 in \eq{PATHINT}. In
that case, the action is purely gluonic (pure gauge theory), and local.
Therefore, using the rules of Metropolis et al.\ we can update each
link independently one by one by some (reversible!) stochastic
modification $U_{\mu}(\vec n)\rightarrow {U'}_{\mu}(\vec n)$, while
only local changes in the action are induced. One `sweep' is performed
if all links are updated once.  By application of the Metropolis rule,
$P(U\rightarrow U')=\min\left[1,\exp({-\Delta S_g})\right]$, detailed
balance is fulfilled, and we are guaranteed to reach the canonical
distribution. Starting from a random configuration, after some
thermalization steps, we can assume hat the generated configurations 
belong to an equilibrium distribution.  Without dynamical
fermions---\ie\ in the quenched approximation---standard
Metropolis shows a complexity $O(V)$, with $V$ being the number of
\dof\ 

However, if we try to use Metropolis for full QCD, the decision
$P(U\rightarrow U')=\min\left[1,\exp({-\Delta
    S_g})\frac{\det(M[U'])}{\det(M[U])}\right]$ would imply the
evaluation of the fermi\-onic determinant for each ${U}_{\mu}(\vec n)$
separately.  A direct computation of the determinant requires $O(V^3)$
operations and therefore, the total computational complexity
would be a number $O(V^4)$.

These implications for the simulation of full QCD with dynamical
fermions have been recognized very early. In a series of successful
steps, the computational complexity could be brought into the range of
quenched simulations\footnote{Take this {\em cum grano salis}.  Two
  $O(V)$ algorithms can extremely differ in the coefficient of $V$.}.
The following table gives an (incomplete) picture of this struggle
towards exact, ergodic, practicable and parallelizable $O(V)$
algorithms for full QCD.
\begin{table}
\caption{Towards exact and  ergodic $O(V)$ algorithms.}
\begin{tabular*}{\textwidth}{@{}|l@{\extracolsep{\fill}}| l | l | l | r|}
\hline
Method                & order & exact & ergodic & year \\
\hline
 Metropolis            & $V^4$ & yes   & yes     & Metropolis et al. 1953\\
Pseudo Fermions       & $V^2$ & no    & yes     & Fucito et al. 1981\\
Gauss Representation  & $V^2$ & yes   & yes     & Petcher, Weingarten 1981\\
Langevin              & $V$   & no    & yes     & Parisi, Wu 1981 \\
Microcanonical        & $V$   & no    & no      & Polonyi at al. 1982\\
Hybrid Molecular Dynamics & $V$ & no  & yes     & Duane 1985 \\
{\sf HMC}                   & $V$   & yes   & yes     & Duane et al. 1987 \\
\hline
Local Bosonic Algorithm & $V$ & no    & yes     & L\"uscher 1994 \\
Exact LBA             & $V$   & yes   & yes     & DeForcrand et
al. 1995\\
5-D Bosonic Algorithm & $V$   & no    & yes     & Slavnov 1996 \\
\hline
\end{tabular*}
\end{table}
A key step was the introduction of the fermionic determinant by a
Gaussian integral.  As a synthesis of several ingredients, HMC is a
mix of Langevin simulation, micro-canonical molecular dynamics,
stochastic Gauss representation of the fermionic determinant, and
Metropolis.

\subsection{Hybrid Monte Carlo: Quenched Case}

For simplicity, I first discuss the quenched approximation, \ie\ 
$\det(M)=\mbox{const.}$ Each sweep of the HMC is composed of two steps:
\begin{enumerate}
\item The gauge field is evolved through phase space by means of
  (micro-canonical) molecular dynamics. To this end, an artificial
  guidance Hamiltonian $\cal H$ is introduced adding the quadratic
  action of momenta to $S_{g}$, ``conjugate'' to the gauge links.  The
  micro-canonical evolution proceeds in the artificial time direction
  as induced by the Hamiltonian.  Choosing random momenta at the begin
  of the trajectory, ergodicity is guaranteed, as it is by the
  stochastic force in the Langevin algorithm. In contrast to Langevin,
  HMC carries out many integration steps between the refreshment of
  the momenta.
\item The equations of motion are chosen to conserve ${\cal H}$. In
  practice, a numerical integration can conserve $\cal H$ only
  approximately.  However, the change $\Delta{\cal H}={\cal
    H}_{f}-{\cal H}_{i}$ is small enough to lead to high acceptances
  of the Metropolis decision---rendering HMC exact, the essential
  improvement of HMC compared to the preceding hybrid-molecular
  dynamics algorihm.
\end{enumerate}
With
\begin{equation}
  {\cal H}= S_{g}[U] +\frac{1}{2}\sum_{\vec n,\mu,color}\mbox{Tr}\,H_{\mu}^{2}(\vec n)
  \quad\mbox{and}\quad Z=\int [dH][dU]e^{-{\cal{H}}},
\end{equation}
expectation values of observables are not altered with respect to
\eq{PATHINT}, if the momenta are chosen from a Gaussian distribution.
A suitable $H$ is found using the fact that $U\in$ SU(3) under the
evolution.  Taylor expansion of $U(\tau+\Delta\tau)$ leads to
$U(\tau)\dot U^{\dagger}(\tau)+ \dot U(\tau)U^{\dagger}(\tau)=0$. This
differential equation is fulfilled choosing the first equation of
motion as
\begin{equation} \dot  U=i H U,\label{FIRST}
\end{equation}
with $H$ represented by the generators of SU(3) and thus being
hermitean and traceless, $H_{\mu}(\vec
n)=\sum_{a=1}^{8}\lambda_{a}{h_{\mu}^{a}(\vec n)}$.  Each component
$h_{\mu}^a$ is a Gaussian distributed random number.  As $\cal H$
should be a constant of motion, $\dot{\cal H}=0$, we get
\begin{eqnarray}
  \dot {\cal H} &=&\sum_{\vec n,\mu}\mbox{Tr}\left\{ H_{\mu}(\vec n)\dot H_{\mu}(\vec n)
    -\frac{\beta}{6}[\dot U_{\mu}(\vec n) V_{\mu}(\vec n)+h.c.]\right\}=0 
\nonumber\\ 
  \dot {\cal H} &=&\sum_{\vec n,\mu}\mbox{Tr}\left\{ H_{\mu}(\vec n) \left[
      \dot H_{\mu}(\vec n)
      -i\frac{\beta}{6}(U_{\mu}(\vec n)V_{\mu}(\vec n)-h.c.)\right]\right\} =0.
\end{eqnarray}
We note that $[]\propto 1$ since $\{ \} $ must be traceless. Since $\dot H$
must stay explicitly traceless under the evolution it follows that $[]=0$.
The second equation of motion reads:
\begin{equation}
  i\dot H(\vec n)= -\frac{\beta}{6}\left\{ U_{\mu}(\vec n) V_{\mu}(\vec n)-h.c.  \right\}.
\end{equation}
The quantities $V_{\mu}(\vec n)$ corresponding to a gluonic force term are
the staples, \ie\ the incomplete plaquettes that arise in the
differentiation, 
\begin{equation}
V_{\mu}(\vec n)=\sum_{\nu\ne\mu}\left\{
\begin{minipage}{2.5cm}
\epsfxsize=2.5cm
\epsfbox{staples.eps}
\end{minipage}
\right\} .
\end{equation}
For exact integration, the Hamiltonian $\cal H$ would be conserved.
However, numerical integration only can stay close to ${\cal H}=const.$
Therefore, one adds a global Metropolis step,
\begin{equation}
  P_{\mbox{\footnotesize acc}}=\min(1,e^{-\Delta\cal H}),
\end{equation} 
to reach a canonical distribution for $\{ U \} $.  As a necessary
condition for detailed balance the integration scheme must lead to a
{\em time reversible} trajectory and fulfill {\em Liouville's
  theorem}, \ie\ preserve the phase-space volume. {\em Symplectic
  integration} is the method of choice. It is stable as far as
energy drifts are concerned.

\subsection{Including Dynamical (Wilson) Fermions}

Dynamical fermions are included in form of a stochastic Gaussian
representation of the fermionic determinant in \eq{PATHINT}.  In order to
ensure convergence of the Gauss integral, the interaction matrix must be
hermitean.  Since the Wilson fermion matrix $M$ is a complex matrix, it
cannot be represented directly. A popular remedy is to consider the two
light quarks {\sf u} and {\sf d} as mass degenerate. With the identity
$\det^{2}(M)=\det(M^{\dagger}M)$ the representation reads
\begin{equation}
  \det(M^{\dagger}M)=\int\left(\prod_{\vec n} [d{\bar\phi_{\vec
        n}}][d\phi_{\vec n}]\right) e^
{-{\phi_{\vec
        n}^{*}(M^{\dagger}M)_{\vec n,\vec m}^{-1}\phi_{\vec m}}}.
\end{equation}
The bosonic field $\phi$ can be related to a vector $R$ of Gaussian random
numbers. In a heat-bath scheme, it is generated using the standard
Muller-Box procedure, and with $\phi=M^{\dagger}R$, we arrive at
$R^{\dagger}R$, the desired starting distribution, equivalent to
$\phi^{*}(M^{\dagger}M)^{-1}\phi= \phi^{*}\,X$.  Adding the fermionic action
to $\cal H$, its time derivative reads:
\begin{eqnarray}
\frac{d S_{f}}{d\tau}&=&
\kappa\sum_{\vec n,\mu}\mbox{Tr}[\dot U_{\mu}(\vec n)F_{\mu}(\vec 
n)+h.c.],\nonumber\\
F_{\mu}(\vec n)&=&
[MX]_{\vec n+\emu}X^{\dagger}_{\vec n}(1+\gamma_{\mu})+
X_{\vec n+\emu}[MX]^{\dagger}_{\vec n}(1-\gamma_{\mu}).
\end{eqnarray}
$F$ is the fermionic force that modifies the second equation of motion to
\begin{equation}
  i\dot H(\vec n)= -\frac{\beta}{6}\left\{ U_{\mu}(\vec n) V_{\mu}(\vec n)
+\kappa\mbox{Tr}\,F_{\mu}(\vec n)
-h.c.  \right\}.
\end{equation}

\subsection{Numerical Integration and Improvements}
The finite time-step integration of the equation of motion must be
reversible and has to conserve the phase-space volume, while it should
deviate little from the surface ${\cal H}=const.$ The leap-frog scheme can
fulfill these requirements.  It consists of a sequence of triades of the
following form:
\begin{eqnarray}
H_{\mu}(\vec n,\tau+\frac{\Delta\tau}{2})&=&
H_{\mu}(\vec n,\tau) + \frac{\Delta\tau}{2}
\dot H_{\mu}(\vec n,\tau)\nonumber\\
U_{\mu}(\vec n,\tau+\Delta\tau)&=&
e^{i\Delta\tau H_{\mu}(\vec n,\tau+\frac{\Delta\tau}{2})}
U_{\mu}(\vec n,\tau)\nonumber\\
H_{\mu}(\vec n,\tau+\Delta\tau)&=&
H_{\mu}(\vec n,\tau+\frac{\Delta\tau}{2})
+ \frac{\Delta\tau}{2}
\dot H_{\mu}(\vec n,\tau+\Delta\tau).
\end{eqnarray}
It can be shown that the leap frog scheme approximates $\cal H$
correctly up to $O(\Delta t^{2})$ for each triade.  As a rule of
thumb, the time step and the number of integration steps, $N_{md}$,
should be chosen such that the length of a trajectory in fictitious
time is $N_{md}\times \Delta\tau\simeq O(1)$ at an acceptance rate
$>70\% $.  It is easy to see from the discrete equations of motion
(EOM) that the phase space volume $[dH][dU]$ is conserved: loosely
speaking, $dU$ is conserved as the first EOM amounts to a rotation in
group space, and from the second EOM follows that $dH'=dH$.  In order
to improve the accuracy of the numerical integration, one can employ
higher order symplectic integrators\footnote{This strategy has been
  used so far only for fine-resolved integration of the gauge fields,
  and coarse resolved integration of the fermions (sparing
  inversions). For small quark masses, this approach can fail,
  however.}.  As the integration part of HMC is not specific for QCD,
higher order integrators could be very useful for other applications
as the Fourier accelerated HMC introduced by A.\ Irb\"ack.

Despite of the reduction of the computational complexity to $O(V)$,
the repeated determination of the large ``vector'' $X$,
$X=(M^{\dagger}M)^{-1}\phi$, renders the simulation of QCD with
dynamical fermions still computationally extremely intensive.  The
size of the vector $X$ is about 1 - 20 $\times 10^{6}$ words.  The
code stays more than 95 \% of execution time in this phase. Since
typical simulations run several months in dedicated mode on  fast
parallel machines, any percent of improvement is welcome.
Traditionally, the system was solved by use of Krylov subspace methods
such as conjugate gradient, minimal residuum or Gauss-Seidel.  In the
last three years, improvements could be achieved by introduction of
the BiCGstab solver \cite{FROMMER94} and by use of novel parallel
preconditioning techniques \cite{FISCHER96} called local-lexicographic
SSOR (symmetric successive over-relaxation). Further improvements have
been achieved through refined educated guessing, where the solution
$X$ of previous steps in molecular dynamics time is fed in to
accelerate the current iteration \cite{BROWER}.  Altogether, a factor
of about 4 up to 8 could be gained by algorithmic research.

\section{Efficiency and Scaling\label{SEC:COMPLEX}}
Apart from purely algorithmic issues, the efficiency of a Monte Carlo
simulation is largely determined by the autocorrelation of the Markov
chain. A significant determination of autocorrelation times of HMC in
realistic full QCD with Wilson fermions could not be carried out until
recently \cite{SESAM97AUTO}.  The length of the trajectory samples in
these simulations was around 5000 (Here, with `trajectory' we denote a
new field configuration at the end of a Monte Carlo decision.).  The
lattice sizes were $16^{3}\times 32$ and $24^{3}\times 40.$

The finite time-series approximation to the
true autocorrelation function for an observable ${\cal O}_t$,
$t=1,\dots,t_{MC}$, is defined as
\begin{eqnarray}
  C^{\cal O}(t)= \frac{\sum\limits_{s=1}^{t_{MC}-t} {\cal O}_s{\cal
      O}_{s+t} - \frac{1}{t_{MC}-t} \left(\sum\limits_{s=1}^{t_{MC}} {\cal
        O}_s \right) ^2}{t_{MC}-t}.
\end{eqnarray}
The definition of corelation in an artificial time is made in analogy
to connected correlation functions in real time.  The {\em integrated}
autocorrelation time is defined as $ \tau^{\cal
  O}_{int}=\frac{1}{2}+\sum_{t'=1}^{t_{MC} \to \infty} \frac{C^{\cal
    O}(t')}{C^{\cal O}(0)}.  $ In equilibrium, $\tau^{\cal O}_{int}$
characterizes the statistical error of the observable ${\cal O}$.

The integrated autocorrelation times have been determined from several
observables, such as the plaquette and the smallest eigenvalue of $M$.
They are smaller than anticipated previously, and their length is
between 10 and 40 trajectories.  Therefore, one can consider
configurations as decorrelated that are separated by $\tau^{\cal
  O}_{int}$ trajectories.

The quality of the data allowed to address the issue of critical
slowing down for HMC, approaching the chiral limit of vanishing {\sf
  u} and {\sf d} quark mass, where the pion correlation length
$\xi_{\pi}=1/m_{\pi}a$ is growing.  The autocorrelation time is
expected to scale with a power of $\xi_{\pi}$,
$\tau=\epsilon\xi_{\pi}^z$, $z$ is called {\em dynamical critical
  exponent}.  As a result the dynamical critical exponent of HMC is
located between $z=1.3$ and 1.8 for local and extended observables,
respectively.

Finally let me try to give a conservative guess of the computational
effort required with HMC for de-correlation.  The pion correlation
length $\xi_{\pi}$ must be limited to
$V^{\frac{1}{4}}/\xi_{\pi}\approx 4$ to avoid finite size effects
as the pion begins to feel the periodic boundary of the lattice.
With $\xi_{\pi}$ fixed, the volume factor goes as $\xi_{\pi}^4$.
Furthermore the compute effort for BiCGstab increases
$\propto\xi^{-2.7}$ \cite{SESAM97AUTO}. In order to keep the
acceptance rate constant, the time step has been reduced (from 0.01 to
0.004) with increasing lattice size ($16^3\times 32$ to $24^3\times
40$), while the number of time steps was increased from 100 to 125.
Surprisingly, the autocorrelation time of the `worst case' observable,
the minimal eigenvalue of $M$, goes down by 30 \% compensating the
increase in acceptance rate cost!  In a conservative estimate, the
total time scales as $ m_{\pi}^{-8}$ to $m_{\pi}^{-8.5}$. As a result,
for Wilson fermions, the magic limit of
$\frac{m_{\pi}}{m_{\rho}}<0.5$, will be in reach on $32^{3}\times
L_{t}$ lattices---on a Teracomputer.

Alternative schemes like the local bosonic algorithm or the
5-dimensional bosonic scheme are by far more memory consuming than
HMC.  Here, a promising new idea might be the Polynomial HMC
\cite{JANSEN}.  The autocorrelation times of these alternative schemes
in realistic simulations are not yet known accurately, however.  In
view of the advantages of HMC mentioned in the introduction, and the
improvements achieved, together with the our new findings as
to its critical dynamics, I reckon HMC to be the method of
choice for future full QCD simulations on Teralcomputers.
\vglue12pt
\noindent{\bf Acknowledgments.}
I thank the members of the SESAM and the T$\chi$L collaborations and A.
Frommer for many useful discussions.

\end{document}